\documentclass[conference,compsoc]{IEEEtran}
\usepackage{upgreek}
\usepackage{hyperref}
\usepackage{url}
\usepackage{graphicx}
\usepackage{booktabs} 
\usepackage{adjustbox}
\usepackage{amsmath}
\usepackage{colortbl} 
\usepackage{xcolor} 
\usepackage[skins]{tcolorbox}
\usepackage[utf8]{inputenc}
\usepackage{dirtytalk}
\usepackage{balance}

\usepackage{amsmath}
\usepackage{mathtools}
\usepackage{amsfonts}
\DeclareMathOperator*{\argmax}{arg\,max}

\ifCLASSOPTIONcompsoc
  \usepackage[nocompress]{cite}
\else
  \usepackage{cite}
\fi

\ifCLASSINFOpdf
  \graphicspath{ {./_img/} }
\else
\fi
\begin{document}
%
% paper title
% Titles are generally capitalized except for words such as a, an, and, as,
% at, but, by, for, in, nor, of, on, or, the, to and up, which are usually
% not capitalized unless they are the first or last word of the title.
% Linebreaks \\ can be used within to get better formatting as desired.
% Do not put math or special symbols in the title.
\title{Towards Understanding and Modeling Empathy for Use in Motivational Design Thinking}

% author names and affiliations
% use a multiple column layout for up to three different
% affiliations

\author{\IEEEauthorblockN{Gloria Washington}
\IEEEauthorblockA{Electrical Engineering and Computer Science \\
Howard University\\
Washington, D.C. 20059\\
Email: gwashington@scs.howard.edu}
\and
\IEEEauthorblockN{Rouzbeh A. Shirvani}
\IEEEauthorblockA{Electrical Engineering and Computer Science \\
Howard University\\
Washington, D.C. 20059\\
Email: rouzbeh.asghari@gmail.com}
}

\maketitle

% As a general rule, do not put math, special symbols or citations
% in the abstract
\begin{abstract}
Design Thinking workshops are used by companies to help generate new ideas for technologies and products by engaging subjects in exercises to understand their users' wants and become more empathetic towards their needs. The "aha moment" experienced during these thought-provoking, step outside the yourself activities occurs when a group of persons iterate over several problems and converge upon a solution that will fit seamlessly everyday life. With the increasing use and cost of Design workshops being offered, it is important that technology be developed that can help identify empathy and its onset in humans.  This position paper presents an approach to modeling empathy using Gaussian mixture models and heart rate and skin conductance.  This paper also presents an updated approach to Design Thinking that helps to ensure participants are thinking outside of their own race's, culture's, or other affiliations' motives. 
\end{abstract}

% no keywords

% For peer review papers, you can put extra information on the cover
% page as needed:
% \ifCLASSOPTIONpeerreview
% \begin{center} \bfseries EDICS Category: 3-BBND \end{center}
% \fi
%
% For peerreview papers, this IEEEtran command inserts a page break and
% creates the second title. It will be ignored for other modes.
\IEEEpeerreviewmaketitle

\section{Introduction and Background}
% no \IEEEPARstart

\subsection{Design Thinking}

When personal computers were initially created they were  difficult to use and learn the functionality, and just non-intuitive. Companies like Microsoft and Apple revolutionized the way we interact with computers by incorporating graphical user interfaces and simplicity into the commands to execute a process. The designers of these interfaces put more focus into user experiences and the satisfaction that users potentially experienced while performing a task with the technology~\cite{brown2009change}. In design thinking exercises, innovative ideas for new technologies are created through activities that challenge individuals to think like a designer.  The main question design thinking tries to answer is how can technology be changed, modified or adapted to better accommodate or address human needs.  Thinking empathetically requires persons to put themselves in another's shoes and experience life as that person. Participants in design thinking workshops often have homework that requires them to observe the world around them (e.g. take public transportation, order food, or observe the repetitive movements in office spaces to access files and resources).  Ideas created as a result should fit seamlessly into a person's life~\cite{brown2009change}. Companies pay thousands of dollars to consultants to conduct Design Thinking Workshops. Design thinking exercises are utilized in several industries including marketing, higher education ~\cite{gardner2017can}, tech companies, and healthcare ~\cite{jones2013design} to create new products or knowledge for bettering an industry~\cite{dym2005engineering}.

\begin{figure}[ht]
\caption{Elements of Design Thinking Exercises}
\centering
\includegraphics[width=0.49 \textwidth] {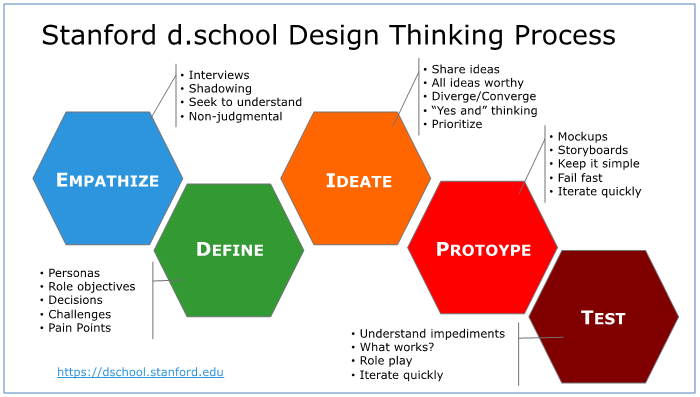}
\end{figure}

Design Thinking has five main steps or modes: 1) Empathize, 2) Define, 3) Ideate, 4) Prototype, and 5) Test. Participants attending Design Thinking workshops often want to improve on an existing idea, process, technology, or artifact that they want to improve.  Or, participants are tasked with creating new product ideas that help to solve an existing  real-world problem. Subjects in the Empathize mode observe and interview users working with the existing artifact or technology. In the case of creating new product ideas,  participants will venture out into society and observe everyday persons encountering the problem they have been tasked to find a solution to. Additionally, they will also hold focus groups interviews to gain more information about the problem and ways humans currently overcome the problem. These observations, shadowing, and interviews help the participants create an assessment of the user's need from only the user's perspective. The Empathize mode is crucial to the creative process in Design Thinking because of perspective-taking. Unfortunately, participants intentionally or unintentionally avoid the mental burden or struggle that comes with experiencing empathy. Empathy is a complex emotion where the intensity of the emotion is displayed uniquely in humans ~\cite{zaki2014empathy}.Affective computing researchers have studied empathy in human-computer interactions; specifically technologies built pertaining to empathy focus more on simulating empathy ~\cite{hall2005achieving} ~\cite{xiao2016computational}. Tools like affective agents for health monitoring applications or social robots ~\cite{paiva2014emotion} have helped to personalize empathetic responses to humans. More research is needed in developing technologies that can learn to recognize empathy and the display of its intensity  in affective interactions. This research examines the complexity of empathy and its intensity through Gaussian Mixture Models (GMMs). GMMs were chosen because the crucial components heart rate and skin conductance ~\cite{metallinou2013tracking} can be built into the model for prediction, the subjective nature of empathy, and for their prior use in emotion recognition   ~\cite{wang2015modeling},~\cite{metallinou2013tracking} . This  is described  in the Methodology.

\begin{figure}[ht]
\caption{Design Thinking Modes. nngroup.com}
\centering
\includegraphics[width=0.45 \textwidth] {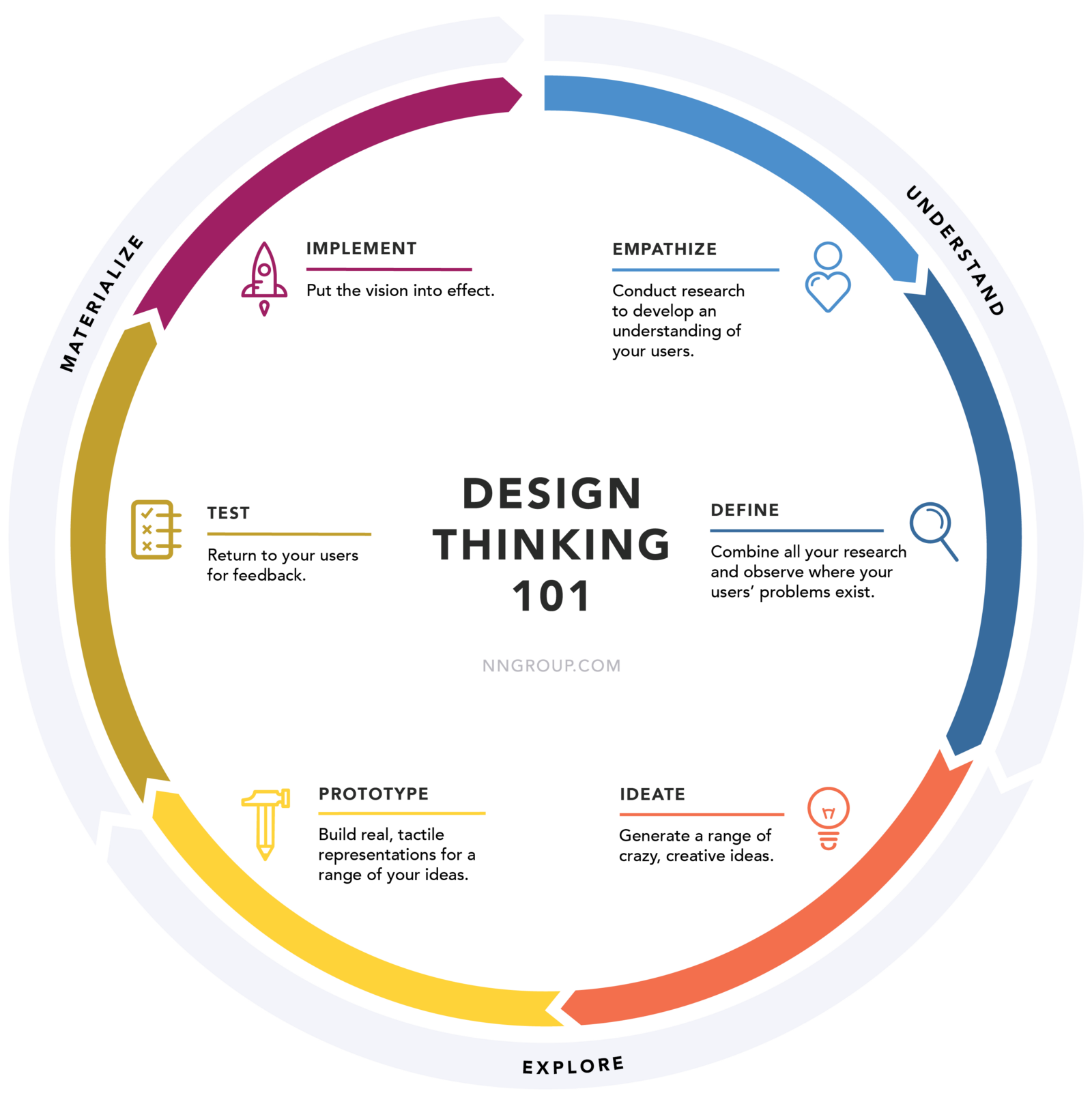}
\end{figure}

This paper is organized by first providing background on the emotion of empathy that is the center of design thinking exercises. This background is focused around the physical aspects of empathy and how it is displayed  by the human body.  Next, a model of empathy is described using a Gaussian Mixture Model technique that encapsulates this complex emotion's components crucial to its detection. Further discussed is the experimental model used to gather data and validate the model.  Lastly, a discussion of the trade-offs and limitations of the technique is provided with an introduction of Motivational Design Thinking approach.

\subsection{What is the Empathy ?}

Empathy is ``the ability to understand and share the feelings of another.`` According to psychology, there are  two categories of empathy; \textit{affective} and \textit{cognitive}. Affective (emotional) empathy refers to the ability to experience and share the emotions of others~\cite{mehrabian1972measure}. Cognitive (perspective-taking) empathy on the other hand,  is the ability to understand the emotions of others~\cite{hogan1969development}. It is not clear whether these two systems are part of a single interacting empathy system or whether they are independent~\cite{shamay2009two}. For this paper, we explore the physical emotional expression of empathy and incorporate affective computing techniques to identify and recognize signs of the emotional reaction in humans. 

\subsubsection{Display of Empathy in Humans}
Empathy is a complex emotion that involves affective arousal, emotion understanding, and emotion regulation ~\cite{decety2010neurodevelopment}.  A human's reaction to another's situation is also called empathetic arousal and it has six categories: circular reaction, classical conditioning, direct association, mimicry, language-mediated association, and role-taking ~\cite{hoffman1984interaction}. It is beyond the scope of this paper to describe each of these modes; however direct association and role-taking are described because they involve events or characteristics that may be encountered in most design thinking activities such as observation of others, watching videos of others engaged in activities, or role-playing. Direct association, ~\cite{humphrey1922multiple}, empathetic arousal occurs when persons observe the facial expressions, voice, posture, or other emotional cues of others in events indicating distress. The observer recalls their distress in a similar past event and experiences the emotion all over again. Role-taking empathetic arousal occurs when a person imagines his/her self in another's place, event, or situation.

Humans as young as two months seem to exhibit some form of empathetic arousal. ~\cite{hoffman1984interaction} observe that babies experience role-taking facial expressions and gestures when interacting with their siblings. However in other studies ~\cite{piaget1932moral},~\cite{shatz1973development}, and ~\cite{imamouglu1975children} empathy occurs between ages 9-12. Males and females display empathy differently with homosexual males and heterosexual females showing lower levels of empathy than heterosexual males~\cite{sergeant2006aggression}. 

Because empathy is an emotional response to someone else's situation, its display in humans is argued to be dependent on one's development in childhood~\cite{hoffman1978toward}.   Persons are said to be highly empathetic if they exhibit a significant increase in skin conductance or heart rate when presented with emotional stimuli designed to produce a response in the observer.   Person's with autism or Asperger's syndrome exhibited low display of empathetic arousal. Additionally, persons with narcissistic personalities or borderline personality disorders are also said to have low physical and emotional arousal when presented with emotional stimuli ~\cite{watson1991narcissism}.

 Most recent studies in psychology show that younger Americans are becoming less empathetic than their parents due to prolonged use of social media and isolation from interpersonal connections outside of technology ~\cite{konrath2011changes},~\cite{zaki2014empathy}, and ~\cite{neumann2011empathy}. ~\cite{zaki2014empathy} suggest that the expression of empathy has changed in younger persons and is related more to internal motivation to approach empathetic arousal if they  experience a positive affect, affiliate with a group engaged in conflict, or desire the social implications of empathy.  However, Zaki et al also mention that  younger Americans avoid showing empathy due to suffering from expected and unexpected consequences, material costs associated with donating money or the mental burden, or interference from competition in hostile negotiations or  when one group believes they are superior to another. Young persons regulate their empathetic arousal through several strategies including opting out of interacting or engaging in activities with suffering persons, dehumanizing persons not affiliated with themselves, and judging or minimizing the amount of suffering experienced by other's not affiliated with themselves.  
 
\subsubsection{Human Physical and Physiological Expression of Empathy}
Previous studies in measuring empathetic arousal rely on self-reporting and physiological responses to emotional stimuli. Skin conductance (palm sweating) or heart rate measures (heart rate variability, ECG, etc.) have been used in prior psychological studies to determine when empathy occurs ~\cite{mehrabian1988emotional}. 

Prior research  has measured empathy from functional magnetic resonance imaging (fMRI) signals indicating changes in the human blood flow in the brain. According to~\cite{keysers2009expanding}, the brain triggers emotional representations while viewing others performing actions or experiencing emotions. The neurons that are responsible for this reaction are called mirror neurons. In, ~\cite{di1992understanding}, researchers observed that specific parts of a monkey's brain increases blood flow when they grasp or hold an object with hand. Additionally, this discovery also led to the discovery that monkey's brains also increase blood blow when a human is performing the same hand movement in front of the monkey. Further, ~\cite{keysers2009expanding, rizzolatti2004mirror}, showed that vicarious activities can be measured for emotions like empathy. Neuroscience studies also showed that different regions of our brain increase brain activity while viewing others being touched, performing actions, or experiencing
\textit{emotions}~\cite{keysers2009expanding, banissy2007mirror}.

In, ~\cite{lamm2007neural}, researchers used functional MRI(fMRI) and demonstrated the level of human empathy could be modulated in a cognitive and motivational process; which results in helping behavior or personal distress. In this study, researchers detected a clear increase in brain activity in subjects who were watching videos of others going through pain or traumatic events. In similar work,~\cite{rizzolatti2005mirror, krolak2003attention} used fMRI imaging and showed that the mirror mechanism localized in parts of brain, like the insula, enables the observer to understand the emotions of others.

Physiological measurement of empathy is another way of quantifying human's empathy in reaction to outside stimuli. Physiological measurement include facial expression, heart rate, galvanic skin response, skin conductance, skin temperature, blood pressure, among others~\cite{neumann2011psychophysiological}. In one study,~\cite{marci2007physiologic} used simultaneous measure of skin conductance between patient and therapist in order to investigate the relationship among physiological concordance, patient-perceived therapist empathy, and social-emotional process during psychotherapy. Their finding suggests that there is significant correlation between skin conductance and social-emotional interactions for both patients and therapists.

~\cite{levenson1992empathy} tested whether empathy between two people is related to a state of shared physiology. In their study, subjects were shown a video in which a married couple (target) are conversing about a disagreement issue. They measured five physiological variables (heart rate, skin conductance, pulse transmission time to the finger, finger pulse amplitude, and somatic activity) from couples as well as subjects who were watching their conversation. Their finding was that there is a relationship between greater physiological linkage and greater ability to rate the negative affect. In other words, they showed that empathy is associated with a state of shared physiology. The more the empathy, the more the similarity between the physiological measurement. In another study,~\cite{westbury2008empathy} presented participants with video clips depicting humans, primates, quadruped mammals and birds in victimized circumstances. As the subject in the video became more similar to human, skin conductance responses (SCR) and subjective empathy increased revealing an empathetic bias towards human stimuli. 

RECOLA~\cite{ringeval2013introducing} is multimodal corpus of spontaneous collaborative and affective interactions in French. While completing a collaborative task, audio, video, ECG and electrodermal activities of participants were recorded. Afterwards, annotators measured emotion continuously on two dimensions: arousal and valence. Using RECOLA,~\cite{feffer2018mixture} trained a personalized shared network based on residual neural networks~\cite{he2016deep} in order to estimate valence and arousal from face images. This work is a notable research in that, it can enhance robots perception about human emotion. In other words, it can empower robot with a very basic level of empathy for the human mood and situation. Video games have been used intensively in recent research in order to trigger people's empathy and measure levels of different emotions during game play~\cite{callejas2017emotion}.

As~\cite{ward2004affective} noted, two problems with physiological measurement of empathy is that 1) the experiments usually occur in tightly controlled laboratory settings and 2) experiments also usually identify human reactions to major events while they should be able to identify reactions to subtle events as well. In design thinking, participants engage in various activities and arise at the best possible solution for a common problem together. This convergence is said to occur when each participant in the activity has reached empathetic thinking about the user. The change in the human body may be very subtle to notice and may be difficult to distinguish from the other emotions displayed in design thinking exercises. Understanding subtle changes in display of empathy in humans is what this research wants to understand and model.  

\section{Methodology}
\subsubsection{Gathering Ground Truth Data - Study I}
Modeling empathy has shown to be a challenging task in affective computing research and most datasets used to train machine learning algorithms on empathy are not publicly or readily available. ~\cite{soleymani2012multimodal} incorporated SVM and modal fusion strategy in order to classify arousal and valence.~\cite{martinez2013learning} tries to address this challenge by 1) feature sets for extracting relevant information from the signal 2) creating affect models that predict affects given feature sets from previous stage. This work is one of the first works that incorporated convolutional neural networks (CNN) for processing physiological signals. 

Inspired by~\cite{martinez2013learning}, ground-truth data will be collected for training the model. Both physiological and self-report measurements will be captured.  In the first experiment,  the control,  subjects will be shown a video a that does not contain a emotional stimuli containing distress.  In another study, the subject will be presented with a video showing a person in some sort of distress (e.g. experiencing a flat tire while its raining outside, a person losing their wallet with their credit cards inside it, or a person spilling coffee on their new pair of white/tan trousers).  Physiological data including ECG, heart rate variability, skin conductance/galvanic skin response will be collected from participants in both the control and the experiment study. 

\subsubsection{Empathy Model}
The data gathered from the Study I (heart rate, heart rate variability, and skin conductance) will be used as input signals , $S_P$ and $S_N$, for training purposes. Rank Margin Error function~\cite{collobert2008unified} will be used in order to train the architecture with the cost function shown in equation~\ref{eq:cost_func}. 

\begin{equation}
    E(S_P,S_N) = max\{0,1-(f(S_P)-f(S_N))\}
    \label{eq:cost_func}
\end{equation}In equation~\ref{eq:cost_func}, $f(S_P)$ and $f(S_N)$ represent the output of our model architecture for preferred and non-preferred samples. Preferred sample is the one that we think contains information about empathy and non-preferred samples contain non or very low information about empathy. This cost function pushes the model toward outputs that are at least 1 measure unit apart from each other. Unlike~\cite{takahashi2012computational}, deep learning approach does not require hand-crafted feature. Rather, we can feed the initial data with little or no pre-processing to the network in order for the learning process to happen. CNNs have been successfully applied to image classification~\cite{krizhevsky2012imagenet} and natural language processing~\cite{collobert2008unified} applications. Convolutional filters in CNN are specifically good at extracting patterns from input signals. This gives us the ability to extract relevant features from our measurements (heart rate, skin conductance, etc.) and as a result have a more accurate model of empathy.

~\cite{wang2015modeling} incorporates probabilistic Gaussian mixture model in order to characterize audio and emotion data. Their generative model is able to recognize the emotion in music as well as retrieving a specific music based on emotion. Inspired by this work, as shown in equation~\ref{eq:seq_measure} we break down a single measurement into a sequence of measurements in which $T$ is the length of the sequence.
\begin{equation}
M = \{m_1, m_2, ..., m_T\}
\label{eq:seq_measure}   
\end{equation}Each measurement is a concatenation of multiple measurement tools (heart rate, heart rate variability, skin conductance, etc.).
\begin{equation}
m_t = \{m_{sc}, m_{hr}, m_{hrv}\}
\label{eq:seq_measure_concat}   
\end{equation}In equation~\ref{eq:seq_measure_concat}, $m_{sc}$, $m_{hr}$, $m_{hrv}$ are measurements for skin conductance, heart rate, and heart rate variability respectively. Each of these measurements will be normalized to unit one. $e$ is the level of empathy that the subject experiences during the experiment. $z$ is the associated latent topic. The corresponding graphical model structure will be as in equation ~\ref{eq:graph_model} in which $e$ is independent of the input $M$.

\begin{equation}
M\to z\to e
\label{eq:graph_model}  
\end{equation}
The distribution of an arbitrary frame x given z is Gaussian. We also assume that the distribution of an arbitrary measurements frame given the latent topic Gaussian.

\begin{equation}
p(m|z=l)=\mathcal{N}(M_l,\Sigma_l)
\label{eq:gauss_latent}  
\end{equation}\begin{equation}
p(m) = \sum_{l}^{L} \pi_l \hspace{1mm} \mathcal{N}(m|M_l,\Sigma_l)
\label{eq:gauss_measure}  
\end{equation}In equation~\ref{eq:gauss_measure}, $\pi_l$, $M_l$, $\Sigma_l$ are model parameters associated with the $l_{th}$ topic. Given the measurement at specific time step, $m_{t}$, the posterior probability of latent topic can be calculated as in equation~\ref{eq:post_latent} 
\begin{equation}
p(z=l|m_t)= \frac{\pi_l \hspace{1mm} \mathcal{N}(m_t|M_l,\Sigma_l)}{\sum_{j=1}^{L}\pi_j \hspace{1mm} \mathcal{N}(m_t|M_j,\Sigma_j)}
\label{eq:post_latent}  
\end{equation}
Equation~\ref{eq:latent_contrib} calculates the latent topic probability of each measurement, assuming that the measurements at all time steps have equal weights.
\begin{equation}
p(z=l|M)= \frac{1}{T}\sum_{t=1}^{T} p(z=l|m_t)
\label{eq:latent_contrib}  
\end{equation}The distribution of the level of empathy, $e$, given the latent topic is Gaussian as in equation~\ref{eq:gauss_label}. $\mu_l$ and $\sigma_l$ are associated with the $l_{th}$ latent topic.

\begin{equation}
p(e|z=l) \sim \mathcal{N} (\mu_l, \sigma_l)
\label{eq:gauss_label}  
\end{equation}Finally equation~\ref{eq:marg_label} provides us with the marginal distribution of the level of empathy given an input measurement.
$$p(e|M)= \sum_{l} p(e|z=l)p(z=l|M) $$
\begin{equation}
= \sum_{l}  \mathcal{N}(\mu_l,\sigma_l)p(z=l|M)
\label{eq:marg_label}  
\end{equation}

According to equation~\ref{eq:marg_label}, $L$ discrete latent topics are used in order to map the measurement into corresponding level of empathy. This can be useful particularly in inferring the relation between measurements and the level of empathy. By looking at equations~\ref{eq:gauss_latent} to~\ref{eq:marg_label}, equation~\ref{eq:param_list} is the list of parameters that we estimate.

\begin{equation}
\Theta \equiv \{\pi_l, M_l, \Sigma_l, \mu_l, \sigma_l\}_{l=1}^L
\label{eq:param_list}  
\end{equation}Maximum likelihood estimation can be incorporated in order to estimate the the parameters of equation~\ref{eq:param_list} from data. In equation~\ref{eq:max_likelihood}, $N$ is the total number of measurements from individual subjects.
\begin{equation}
\hat{\Theta}={\argmax_\theta}\sum_{i=1}^{N}log\hspace{1mm}p(Y^{\small(i\small)}|M^{\small(i\small)},\Theta)
\label{eq:max_likelihood}  
\end{equation}The overall algorithm for learning the model parameters in equation~\ref{eq:param_list} can be described as follows:
\begin{itemize}
  \item Dividing each subject's measurements into different time steps as in equation~\ref{eq:seq_measure}. 
  \item Use equation~\ref{eq:post_latent} and~\ref{eq:latent_contrib} in order to calculate $p(z|M)$
  \item Use equation~\ref{eq:marg_label} in order to calculate $p(e|M)$ which will give us the distribution of empathy. In other words a weighted combination of $\{\mathcal{N}(\mu_l,\sigma_l)\}_{l=1}^L$ as in equation~\ref{eq:marg_label} will provide us with the distribution of empathy. Finally, given the model parameters $\{\pi_l, M_l, \Sigma_l, \mu_l, \sigma_l\}_{l=1}^L$, we can predict the level of empathy corresponding to each measurement.
\end{itemize}
% 1. https://opus.lib.uts.edu.au/bitstream/10453/4108/1/2007000499.pdf

The model described above will be tested on a test set of data from Study I.  Predictions from the model will be compared against the self-report measures that the participants provided to verify the results accuracy of the model. Additionally, this model will be compared against other machine learning algorithms to determine the performance of the model against other techniques. 

\section{Experiment Design}
\subsubsection{Design Thinking Study}
Public transportation is a huge headache for residents within the Washington, DC area because of  traffic related to tourism, national security concerns, and persons heading to and from jobs inside the public and private sectors. Additionally, transportation in DC is more of an inconvenience due fluctuations of persons that call the District home only half the year.  To further test the model described in this work, participants in the Washington, DC area will be asked to volunteer to engage in a design thinking exercise where they create the next best mode of public transportation. The study will call for six participants, 3 female and 3 male, to wear a watch-like device that gathers heart rate and galvanic skin response measures.  Participants from different races and socio-economic backgrounds will be included in the study.

Participants before starting the study will answer a pre-survey using the Positive and Negative Affect Schedule ~\cite{thompson2007development} to  determine the mood of the individuals before they participate in any of the activities and to help the researchers baseline for the affect of the persons involved in the study.  Additionally, participants will fill-out a Meyers-Briggs personality assessment  to determine if they have traits outlined in the Design Thinker's Personality Profile ~\cite{brown2009change}.  After key activities in the study, subjects will be asked to self-report their emotional levels using the Empathy Quotient (EQ) short questionnaire ~\cite{wakabayashi2006development}. After completion, participants will answer the PANAS and a questionnaire about their experience in the study.

\section{Discussion}
Innovation thrives through diversity: diversity of thought, experience, opinion, etc. Nonetheless, some people involved in design thinking exercises intentionally or unintentionally avoid the most important part of the activity: experiencing empathy. Because of this, there isn't a guarantee that these often costly workshops will yield the best ideas. As mentioned, empathy is decreasing in younger populations exposed to long-term use of social media platforms like Facebook, Twitter, etc because of the lack of true social connections developed in the real-word.  Tech company employee bases are growing younger and younger with the the average age of a software developer being 31 ~\cite{eadicicco_2015}.  Persons that engage in design thinking workshops are usually mid to senior-level product designers or developers, product managers, entrepreneurs, and other professionals tasked with creating new ideas for a company. These individuals are on average 35+ and can possibly remember a time when social connections were established and fostered in the offline world. 

To ensure that participants in the Design Thinking Study are not avoiding or emotionally regulating their responses to the activities in the exercises, the participants will be briefed on the malleable theory of empathy ~\cite{zaki2014empathy}. This theory has shown if persons are informed of 1) the drawbacks to feeling empathy like painfulness to remembering past experiences and 2) being open to listening to conflicting opinions and engaging in conversations with persons outside of their affiliation; their attitude towards experiencing empathy can be changed.  In addition to the limitations surrounding empathy display already discussed; limitations also exist between different racial groups. Empathetic arousal decreases for persons with different racial backgrounds ~\cite{avenanti2010racial} ~\cite{finlay2000improving} with the differences mainly attributed to culturally acquired prejudice. We assume that participants in the Design Thinking Study will come into the study with cultural prejudices. Malleable empathy theory has also shown to help persons not within the same racial group ward off empathy avoidance.

\begin{figure}[ht]
\caption{Elements of Motivational Design Thinking Exercises}
\centering
\includegraphics[width=0.52 \textwidth] {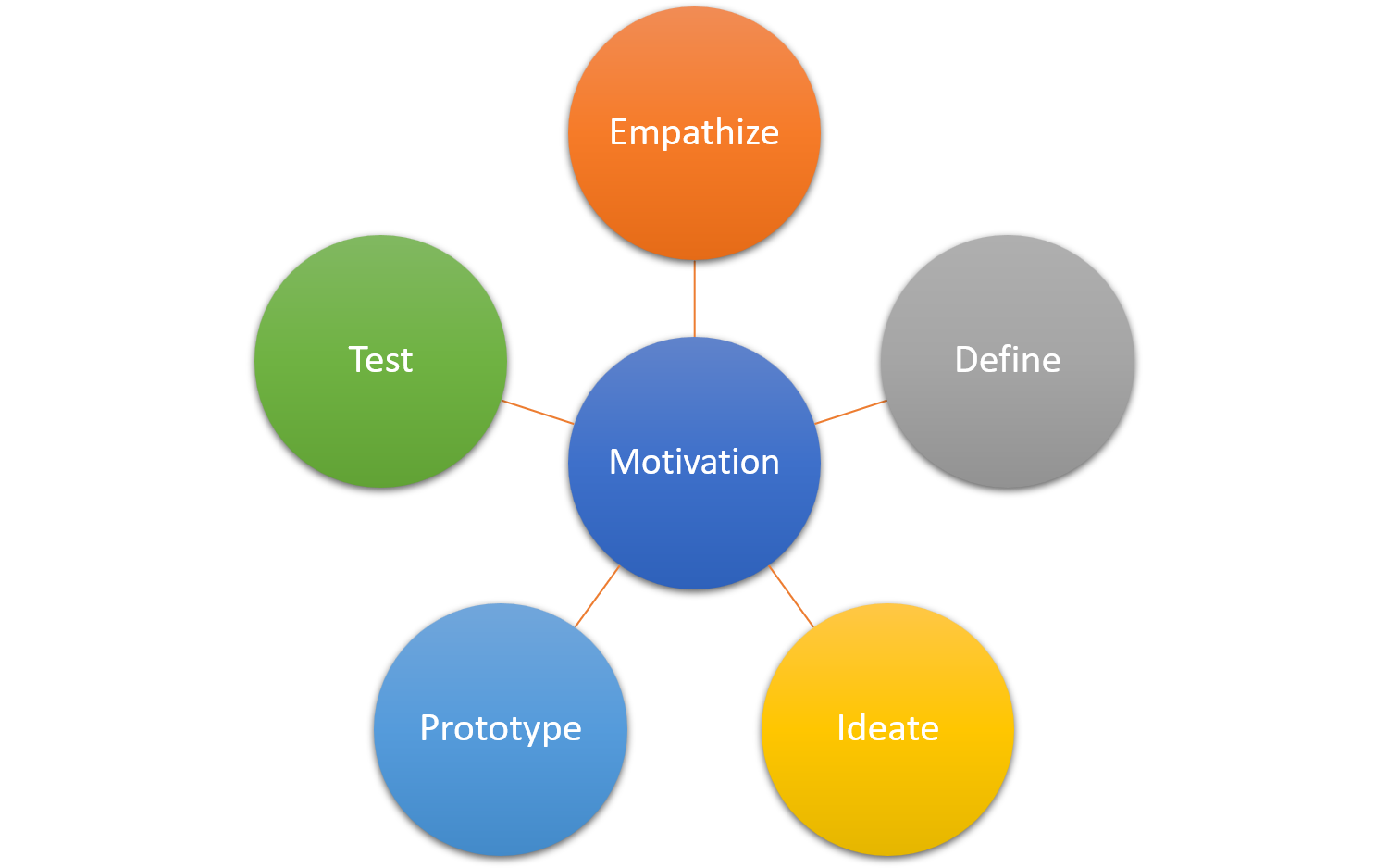}
\end{figure}

\subsection{Motivational Design Thinking}
The limitations of design thinking directly relate to the persons involved in the activities. To offset these limitations, motivational design thinking is proposed to help participants engaged in design thinking experience empathy outside of their group. It is predicated off of the motivational theory of empathy that helps refocus the motives of the participants with strategies used to increase empathy. These interventions, borrowed from psychological  studies ~\cite{zaki2014empathy} include:
\begin{itemize}
\item Perspective-taking exercises 
\item Training persons on observation techniques that help identify different emotions through facial expressions
\item Mindful thinking or compassionate meditation
\item Narrative reading
\end{itemize}
Perspective-taking exercises and training on facial expressions interventions should occur in the Empathize mode because participants  engage in interviews and shadow their users. Additionally, facial expression training may also occur in the Test mode as participants engage in role-playing or testing of their prototype solutions to determine what works for their users.  Narrative reading interventions should occur in the Define mode as participants develop Personas used to describe the habits and priorities of the user. Motivational interventions will be employed in the Design Thinking Study.

\section{Conclusion}
This paper outlines a computational model solution for  use in prediction of display of empathy often experienced in design thinking exercises. Understanding the impact of design thinking exercises has not been fully quantified even though some studies suggest these exercises have helped to reduce implicit and cognitive bias ~\cite{liedtka2015perspective}, redesigning the way teams work together ~\cite{liedtka2014innovative}, and improving customer satisfaction ~\cite{schmiedgen2016measuring}. Silicon Valley and tech companies outside of the valley are investing thousands of dollars in these Design Thinking Workshops like those held at Standord's D School and IDEO. This work will help to quantify if the participants in these activities are actually experiencing an increase in empathetic arousal due to participating in the modes of design thinking. Additionally, little is known about how empathy is expressed in the human body when creating new innovations; however through the Design Thinking Study described in this paper; we hope to shed more light on  this process through measuring empathetic arousal. The model described in this paper will be validated against other machine learning techniques like CNNs.  

This paper also describes a Motivational Design Thinking that infuses intervention strategies to increase empathy in persons participating in design activities. Motivational Design Thinking has not been validated, but this study will help to analyze if the approach for helping humans change their perspective and think outside of their race, cultural affiliations, and socio-economic status helps generate better ideas.  We hope this paper will help spark conversations between psychologists and affective computing researchers to develop technology that will help further understand the complexity of empathy exhibited in humans.

% conference papers do not normally have an appendix

% use section* for acknowledgment
% \ifCLASSOPTIONcompsoc
%   % The Computer Society usually uses the plural form
%   \section*{Acknowledgments}
% \else
%   % regular IEEE prefers the singular form
%   \section*{Acknowledgment}
% \fi

% The authors would like to thank...

\balance
\bibliography{ref}
\bibliographystyle{IEEEtran}

% that's all folks
\end{document}